\documentstyle[12pt]{article}
\font\tenrm=cmr10

\def\th{\theta}

\def\ka{\kappa}

\def\De{\Delta}

\def\cv{{\cal V}}
\def\fr#1#2{{{#1} \over {#2}}}

\def\expect#1{\langle{#1}\rangle}

\def\frac#1#2{{\textstyle{{#1}\over {#2}}}}

\def\lsim{\mathrel{\rlap{\lower4pt\hbox{\hskip1pt$\sim$}}
    \raise1pt\hbox{$<$}}}
\def\gsim{\mathrel{\rlap{\lower4pt\hbox{\hskip1pt$\sim$}}
    \raise1pt\hbox{$>$}}}
\def\sqr#1#2{{\vcenter{\vbox{\hrule height.#2pt
         \hbox{\vrule width.#2pt height#1pt \kern#1pt
         \vrule width.#2pt}
         \hrule height.#2pt}}}}

\newcommand{\beq}{\begin{equation}}
\newcommand{\eeq}{\end{equation}}
\newcommand{\bea}{\begin{eqnarray}}
\newcommand{\eea}{\end{eqnarray}}
\newcommand{\rf}[1]{(\ref{#1})}

\renewenvironment{thebibliography}[1]
 { \rm
   \begin{list}{\arabic{enumi}.}
    {\usecounter{enumi} \setlength{\parsep}{0pt}
     \setlength{\itemsep}{3pt} \settowidth{\labelwidth}{#1.}
     \sloppy
    }}{\end{list}}

\def\theequation{\thesection.\arabic{equation}}
\def\@eqnnum{{\rm (\theequation)}}

\newcommand{\bref}[1]{(\ref{#1})}
\newcommand{\ct}[1]{\cite{#1}}
\def\lrb#1{\left( { #1 } \right)}
\def\ie{i.e.}
\def\c2th{\cos{ 2\theta }}
\def\s2th{\sin{ 2\theta }}
\def\dds#1{{{d #1} \over {ds}}}
\def\secttit#1{\vglue 0.6cm{\bf\large\noindent{#1}}\vglue 0.4cm}
\def\eq{Eq.\ }
\def\eqs{Eqs.\ }

\begin{document}
\titlepage

\begin{flushright}
{IUHET 291\\}
{MPI--PhT/94--80\\}
{CCNY--HEP--94/8\\}
{hep-ph/9506262\\}
{December 1994\\}
\end{flushright}
\vglue 1cm

\begin{center}
{
{\Large \bf Self-Maintained Coherent Oscillations\\
in Dense Neutrino Gases
\\}
\vglue 1.0cm
{V. Alan Kosteleck\'y$^{a}$ and Stuart Samuel$^{b*}$\\}
\bigskip
{\it $^a$Physics Department\\}
{\it Indiana University\\}
{\it Bloomington, IN 47405, U.S.A.\\}
\vglue 0.3cm
\bigskip

{\it $^b$Max-Planck-Institut f\"ur Physik\\}
{\it Werner-Heisenberg-Institut\\}
{\it F\"ohringer Ring 6\\}
{\it 80805 Munich, Germany\\}

\vglue 0.8cm
}

\vglue 0.3cm

{\bf Abstract}
\end{center}
{\rightskip=3pc\leftskip=3pc\noindent
We present analytical solutions to
the nonlinear equations describing
the behavior of a gas of neutrinos
with two flavors.
Self-maintained coherent flavor oscillations
are shown to occur when the gas density
exceeds a critical value
determined by the neutrino masses
and the mean neutrino energy in the gas.
Similar oscillations may have occurred
in the early Universe.

}

\vskip 0.5truein
\centerline{\it To appear in Physical Review D, July 1995}
\vfill

\textwidth 6.5truein
\hrule width 5.cm
\vskip 0.3truecm
{\tenrm{
\noindent
$^*$Permanent address: Physics Department,
City College of New York,
New York, NY 10031, USA.\\
\hspace*{0.2cm}E-mail: samuel@scisun.sci.ccny.cuny.edu\\}}

\newpage

\baselineskip=20pt

{\bf\large\noindent I.\ Introduction}
\vglue 0.4cm
\setcounter{section}{1}
\setcounter{equation}{0}

The properties of neutrinos
passing through matter have attracted
much attention in the past few years.
The reason is
that the large electron-lepton number of matter
results in flavor-dependent neutrino propagation
and,
at suitable densities,
enhanced neutrino flavor oscillation
\cite{msw}.
The effect can be attributed primarily to
neutrino-electron forward scattering
through $W^\pm$ exchange.
Among the consequences is a possible resolution of
the solar neutrino problem
\cite{bahcall}.

Another situation with lepton imbalance is
the early Universe
\cite{kt}
at a temperature below 1 MeV,
where neutrino propagation
is affected by the excess of electrons over positrons.
Some implications for the early Universe
of neutrino oscillations enhanced by $W^{\pm}$ exchange
have been considered in
\cite{lpst}.
In this scenario,
however,
the self-interactions of neutrinos through $Z^0$ exchange
cannot be neglected
because the neutrino density is relatively high.
Indeed, numerical simulations of the early Universe
show that neutrino properties can be significantly modified
by these effects
\cite{kps93,ks93,ks94}.
In the parameter region of
\cite{ks93},
there is even
a novel neutrino-flavor conversion mechanism
that is different from the MSW effect.

The neutrino self-interactions are nonlinear.
In a general context,
they must be studied numerically.
Moreover,
in the early Universe,
neutrino behavior is controlled by a combination
of factors.
In addition to the electron-positron imbalance
and the presence of the neutrino gas itself,
these include other effects such as
the expansion rate of the Universe.
In this paper,
we eliminate these additional complications
by considering a simplified situation
consisting of a homogeneous gas
of self-interacting neutrinos in a box
of fixed volume $\cv$,
with no other leptons present.
Our goal is to obtain
results that are analytical but that nonetheless
describe nonlinear features of neutrino behavior.

We consider here two situations:
the pure neutrino gas (Sec.\ 3),
and a gas containing both neutrinos and antineutrinos
(Sec.\ 4).
Throughout the paper,
we assume that hard-scattering processes
are negligible compared to forward scattering.
This is valid provided the energy $E$ of the neutrino
satisfies the condition
$G_F E^2 / {\lrb{\hbar c}}^3 \ll 1$.
Forward scattering corresponds to phase-interference effects
and hence to neutrino oscillations.
Under these conditions,
neutrinos are neither created nor destroyed
but are simply transformed from one flavor to another.

For simplicity in what follows,
we restrict ourselves to oscillations between
electron and muon neutrinos.
For vacuum oscillations,
relevant parameters are
the vacuum mixing angle $\th$ and
the mass-squared difference $\De = \lrb{m_2^2 - m_1^2}c^4$.
The effective hamiltonian describing
free neutrino propagation is diagonal
in the mass-eigenstate basis.
However,
weak-interaction processes produce and destroy
flavor-eigenstate neutrinos.
The mass-eigenstate neutrinos $\nu_1$ and $\nu_2$
are related to
the left-handed flavor-eigenstate neutrinos
$\nu_{e L}$ and $\nu_{\mu L}$
by
\beq
 \nu_1 =
  \nu_{e L} \cos{\th} - \nu_{\mu L} \sin{\th}
\quad , \qquad
 \nu_2 =
  \nu_{e L} \sin{\th} + \nu_{\mu L} \cos{\th}
\quad ,
\label{nu1}
\eeq
where $m_1$ and $m_2$ are the masses of $\nu_1$ and $\nu_2$,
respectively.

The vacuum-oscillation period $T_{\De}$ is given by
\beq
 T_{\De} = \fr{4 \pi E \hbar}{\De}
\quad .
\label{TDelta}
\eeq
For a gas of $N_{\nu}$ neutrinos
with a finite energy spread,
neutrinos oscillate with various periods.
In the limit of negligible interactions,
a neutrino (or mixed neutrino-antineutrino)
gas with large $N_\nu$ exhibits oscillation decoherence,
i.e.,
the net neutrino-flavor content
is time independent at late times.
This constant asymptotic behavior
is independent of initial conditions.
For example,
suppose one begins with a gas of electron neutrinos.
Some will convert into muon neutrinos
so that time-varying behavior occurs initially.
The flavor content of an individual neutrino
at time $t$ is
$
1 - \sin^22\th
\lrb{ 1 - \cos{\lrb{ {2 \pi t }/{T_\De}}} }
$
for $\nu_e$
and
$
\sin^22\th
\lrb{ 1 - \cos{\lrb{ {2 \pi t }/{T_\De}}} }
$
for $\nu_{\mu}$.
The summation over cosine functions
with various periods $T_\De$
leads asymptotically to a constant function of $t$
if sufficiently many terms are present, \ie,
if $N_\nu$ is sufficiently large.
The ratio of electron neutrinos to muon neutrinos
in the limit $t \to \infty$ becomes the constant factor
$(1 - \sin^22\th)/\sin^22\th $.

Vacuum behavior dominates
provided the neutrino gas is sufficiently dilute,
so that interactions due to $Z^0$ exchange
remain unimportant.
The dominance is controlled by
a dimensionless parameter $\ka$,
defined by
$$
  \ka =
   \fr{\De}{2 \sqrt{2} G_F E n_\nu }
\quad ,
$$
where $n_\nu = N_\nu / {\cal V}$
is the neutrino density.
This parameter can be written
as the ratio $T_{\nu}/T_{\De}$,
where
\beq
 T_{\nu} =
    \fr{2 \pi \hbar}{\sqrt{2} G_F n_\nu }
\quad
\label{Tnu}
\eeq
is the time scale
associated with neutrino interactions.
When the neutrino density is low,
$\ka$ is large.
Neutrino-neutrino forward scattering occurs infrequently
compared to a vacuum oscillation period,
and the behavior is similar to a non-interacting gas.
This region of parameter space is characterized
by decoherence.

In contrast,
when the neutrino density is large
so that $\ka \ll 1$,
neutrino interactions are important.
Many neutrino-neutrino interactions occur
during a vacuum-oscillation period.
Numerical simulations
for the pure neutrino gas
\ct{samuel93}
reveal the existence in this parameter region
of a collective mode of the nonlinear dynamics
in which the behaviors of individual neutrinos
are correlated.
Significant numbers of neutrinos oscillate in unison.
We refer to this
counterintuitive behavior as self-maintained coherence.
Self-maintained coherence is also seen in numerical
simulations of a gas of neutrinos and antineutrinos
\ct{kps93,ks93,ks94}.
A system consisting initially of electron neutrinos
does not decohere.
Instead, oscillatory behavior is observed,
even at late times.
A primary goal of this paper
is to obtain an analytical description
of self-maintained coherence
for $\ka \ll 1$.

If $m_2 < m_1$ so that $\De < 0$
(or, alternatively, if $\De > 0$ and $\th > \pi/2$),
then there is a large region
in the $\De$-$\th$ parameter space
for which self-maintained coherence emerges
for neutrino oscillations in the early Universe
\ct{ks93}.
The behavior begins smoothly,
but, due to the expansion of the Universe
and the varying electron and positron densities,
coherent oscillations emerge
at around $100$ seconds after the big bang.
Self-maintained coherent oscillations
may thus have played a role in early-Universe physics.

Throughout the rest of this paper,
we work in units with $\hbar = c = 1$.

\secttit{II.\ Background Material}
\setcounter{section}{2}
\setcounter{equation}{0}

A single relativistic neutrino
oscillating in vacuum obeys the equation
\beq
 i \fr{d \nu}{dt} =  H \nu
 \quad ,
\label{Heff}
 \eeq
where $\nu (t)$ is
the two-component flavor wave function
\beq
  \nu  =
\lrb{\matrix{
     \nu_e   \cr
     \nu_\mu \cr}}
\quad
\label{nu}
\eeq
with $\nu_e^{*} \nu_e + \nu_\mu^{*} \nu_\mu = 1 $,
and where the effective hamiltonian $H$ is given by
\beq
H=
   \fr{m_1^2 + m_2^2}{4 E}
\lrb{\matrix{
  1&0\cr
  0&1\cr}}
       +
   \fr{\De}{4 E}
\lrb{\matrix{
   -\c2th&\s2th \cr
    \s2th&\c2th \cr}}
\quad .
\label{dnudt}
\eeq
The probability for the particle
to be an electron neutrino is $\nu_e^{*} \nu_e$,
while that to be a muon neutrino is
$\nu_\mu^{*} \nu_\mu$.

A convenient and standard vector reformulation
of the above equations exists
\ct{kim}.
It is useful both for visualization
and for numerical simulation of oscillations.
Define the vector
\beq
 \vec v \equiv
          \lrb{\nu_e^{*}\nu_e - \nu_\mu^{*}\nu_\mu,
 2 {\rm Re} ( {\nu_e^{*}\nu_\mu}),
 2 {\rm Im} ( {\nu_e^{*}\nu_\mu}) }
\quad .
\label{vecv}
\eeq
Then,
the neutrino oscillation equation
\bref{dnudt}
is equivalent to that governing
a particle of unit mass and charge
moving in a magnetic field $\vec B$
given by
\beq
  \vec B  = \fr{\vec \De}{2E}
\quad ,
\label{vecB}
\eeq
where
\beq
  \vec \De \equiv
     \De \lrb{ \c2th, -\s2th, 0 }
\quad .
\label{vecDelta}
\eeq
For an antineutrino,
\eqs\bref{dnudt}--\bref{vecDelta} hold
if neutrino wave functions
are replaced by antineutrino wave functions, \ie,
$\nu \to \bar \nu$.
We denote the corresponding vector
for an antineutrino by $\vec w$.
Throughout this work,
we use the expression
``magnetic field''
to refer to an effective magnetic field,
as opposed to a physical one.

For a gas,
there is a vector $\vec v^j$ for the $j$th neutrino
and a vector $\vec w^k$ for the $k$th antineutrino.
The equations governing
the self-interacting gas become
\ct{kps93,sr93}
\beq
 {{d\vec v^j} \over {dt}} = \vec v^j \times \vec B_v^j
\quad
\label{dvecvjdt}
\eeq
for vectors associated with neutrinos,
and
\beq
 {{d\vec w^k} \over {dt}} = \vec w^k \times \vec B_w^k
\quad
\label{dvecwkdt}
\eeq
for antineutrinos.
Here,
the magnetic fields
$\vec B^j_v$ and $\vec B^k_w$
are given by
\beq
  \vec B_v^j =
   {{\vec \De } \over {2E^j}} -
    {\vec V_{\nu \nu}}
\ , \quad \quad
 \vec B_w^k =
   {{\vec \De } \over {2\overline E^k}} +
    {\vec V^{*}_{\nu \nu}}
\quad ,
\label{vecBvj}
\eeq
where the energy of the $k$th antineutrino is
denoted $\overline E^k$.
An asterisk on a vector indicates
a change in sign of the third component.
The vacuum contributions to the magnetic fields are
the terms in \eq\bref{vecBvj}
dependent on $\vec \De$,
given in
\eq\bref{vecDelta}.
The potential $\vec V_{\nu \nu}$
is generated by $Z^0$ exchange
and is given by
\beq
    \vec V_{\nu \nu}   =
   {\fr{\sqrt 2 G_F}{\cal V}}
     \lrb{ \expect{\vec v} - \expect{\vec w^*}}
\quad ,
\label{vecVnunu}
\eeq
where
$
  G_F \simeq 1.17 \times
10^{-11} \, {\rm MeV}^{-2}
$
is the Fermi coupling constant,
and
\beq
  \expect{\vec v} = \sum\limits_j {\vec v^j}
\ , \quad \quad
  \expect{\vec w} = \sum\limits_k {\vec w^k}
\quad .
\label{averagevecv}
\eeq

In the absence of ${\vec V_{\nu \nu}}$,
the first-order differential equations
in \bref{dvecvjdt} and \bref{dvecwkdt}
decouple and are linear.
The system is then solvable
and the solution corresponds
to a non-interacting gas
in which each neutrino undergoes
vacuum oscillatory behavior.
When ${\vec V_{\nu \nu}}$ is present,
the equations
in \bref{dvecvjdt} and \bref{dvecwkdt}
are both coupled and nonlinear.
For this reason,
we refer to ${\vec V_{\nu \nu}}$
as the nonlinear term.
This nonlinearity
leads to interesting effects.

Individual neutrinos and antineutrinos are
neither created nor destroyed
under our assumptions.
The equations expressing this,
\beq
  {{d \lrb{ \vec v^j \cdot \vec v^j } } \over {dt}} = 0
\ , \quad \quad
  {{d \lrb{ \vec w^k \cdot \vec w^k } } \over {dt}} = 0
\quad ,
\label{dvecvjdotvecvjdt}
\eeq
follow from
\eqs\bref{dvecvjdt} and \bref{dvecwkdt}.

Different normalizations of $\vec v^j$ are possible.
Above,
we have chosen $\vec v^j \cdot \vec v^j = 1$
along with the interpretation that
the index $j$ labels individual neutrinos.
For this case,
the index $j$ ranges
from $1$ to $N_{\nu}$,
where $N_{\nu}$ is the total number of neutrinos.
A second normalization convention follows
from noting that neutrinos with the same energy obey
the same oscillation equation.
One can therefore perform a sum
over all vectors of the same energy.
With this second normalization,
$| \vec v^j | $
represents the number of neutrinos of energy $E^j$.
The index $j$ then ranges over the possible energy values.
Similar normalization choices exist for antineutrinos.

For both the above normalization conventions,
the total number of neutrinos $N_{\nu}$
and antineutrinos $N_{\bar \nu}$ is given by
\beq
 N_{\nu} =
 \sum_j | \vec v^j |
\ , \quad \quad
 N_{\bar \nu} =
 \sum_k | \vec w^k |
\quad .
\label{sumjvj}
\eeq
The neutrino and antineutrino densities
$n_{\nu}$ and $n_{\bar \nu}$
can then be obtained
by dividing{\footnote{
Both normalization schemes discussed here
can also be modified
by multiplying neutrino vectors by ${1}/{{\cal V}}$.
In this situation,
neutrino vectors become densities,
and
${\cal V}$ in the nonlinear term $\vec V_{\nu \nu}$
given in \eq\bref{vecVnunu}
must be replaced by $1$.}}
by ${\cal V}$,
that is,
$n_{\nu} = N_{\nu}/{\cal V}$,
$n_{\bar \nu} = N_{\bar \nu}/{\cal V}$.

For numerical purposes
the second normalization scheme
is more useful.
For the mathematical treatment
in the current work,
we use the first convention
with $\vec v^j \cdot \vec v^j = 1$
and
$\vec w^k \cdot \vec w^k = 1$.

\secttit{III.\ The Pure Neutrino Gas}
\setcounter{section}{3}
\setcounter{equation}{0}

In this section,
we analyze self-maintained coherence
for a pure neutrino gas.
For this system,
$N_{\bar \nu} = 0$
and the contribution to the neutrino potential becomes
\beq
  {\vec V_{\nu \nu}}  =
   {\fr{\sqrt 2 G_F}{\cal V}} \expect{\vec v}
\quad .
\label{3vecVnunu}
\eeq
For definiteness,
we consider the situation
in which
$N_\nu$ electron neutrinos
are placed in the box at time $t=0$,
so that
the initial conditions are
\beq
\vec v^j ( 0 ) = \lrb{ 1, 0, 0 }
\quad .
\label{3vecvj0}
\eeq
At $t=0$,
the ratio $\ka^j$ of the vacuum term
to the neutrino-neutrino term
for the $j$th neutrino is
\beq
 \ka^j = \fr{\De}{ 2 \sqrt{2} G_F E^j n_{\nu} }
\quad .
\label{3kappaj}
\eeq
When $\ka^j \ll 1$,
the vacuum term is dominated by
the neutrino-neutrino interaction term,
and self-maintained coherence appears
in computer simulations
\ct{samuel93}.
Neglecting the non-linear term,
neutrinos with larger energies
oscillate slower
and neutrinos with smaller energies
oscillate faster.
However,
a large ${\vec V_{\nu \nu}}$ term
boosts slow neutrinos
and retards fast neutrinos.

To obtain an analytical solution,
we can take advantage of a feature
of the motion called alignment
\ct{ks94}:
numerical simulation shows that
vectors in the nonlinear system
point in a common direction
when the $\ka^j$ are small.
Alignment of the $j$th neutrino
implies that the approximation
\beq
  \vec v^j (t) \approx
  \fr{\expect{\vec v}}{N_\nu} \equiv
   \vec r_{v} (t)
\quad
\label{3vjt}
\eeq
is good.
Here,
$\vec r_{v} (t) $
is the average neutrino vector.
This feature suggests we seek an analytical solution
for $\vec r_v(t)$.

An equation for $\vec r_{v} (t) $
can be obtained
by summing over $j$
in \eq\bref{dvecvjdt}
and using
\eqs\bref{vecBvj}, \bref{averagevecv} and \bref{3vecVnunu}:
\beq
   {{d\vec r_v } \over {dt}} =
  \vec r_v \times \fr{\vec \De}{2 E_0}
\quad ,
\label{3dvecrvdt}
\eeq
where the average inverse energy
$1/E_0$
is defined by
\beq
  \fr{1}{E_0} \equiv
  \fr 1 {N_\nu}  \sum_j \fr{1}{E^j}
\quad .
\label{31E0}
\eeq
Equation \bref{3dvecrvdt}
shows that the self-maintained coherence in the
pure neutrino gas is formally equivalent
to the oscillation of a single neutrino in vacuum.
Hence,
the average neutrino vector
undergoes vacuum oscillations
with an effective energy $E_0$.

It is useful for later purposes,
when the neutrino-antineutrino gas is considered,
to display the solution
of \eq\bref{3dvecrvdt}.
The initial conditions for $\vec r_v$ are
\beq
   \vec r_v (0) = \lrb{ 1, 0, 0 }
\quad .
\label{3vecrv0}
\eeq
Equation \bref{3dvecrvdt},
when written in components,
is
$$
   {{dr_{v1} } \over {dt}} =
      \lrb{ \fr{\De}{2 E_0} \s2th } r_{v3}
\quad ,
$$
$$
   {{dr_{v2} } \over {dt}} =
      \lrb{ \fr{\De}{2 E_0} \c2th } r_{v3}
\quad ,
$$
\beq
   {{dr_{v3} } \over {dt}} =
    - \fr{\De}{2 E_0}
      \lrb{ r_{v1} \s2th + r_{v2} \c2th }
\quad .
\label{3drv1dt}
\eeq
These equations simplify
in the vacuum-mass-eigenstate basis
denoted by $\vec R (t)$
and given by
\beq
  R_1 \equiv r_1 \c2th - r_2 \s2th
\quad , \quad
  R_2 \equiv r_1 \s2th + r_2 \c2th
\quad , \quad
  R_3 \equiv r_3
\quad .
\label{3R1}
\eeq
In this basis,
the equations resemble those
in \eq\bref{3drv1dt} with $\th = 0$.
Thus, $R_1 (t)$ is a constant.
The equations for $R_2$ and $R_3$ combine to give
a harmonic-oscillator system.
Incorporating the initial conditions
\bref{3vecrv0},
we find
$$
  R_1 (t) =  \c2th
\quad ,
$$
$$
  R_2 (t) =   \s2th \cos{\lrb{ \fr{\De}{2 E_0} t }}
\quad ,
$$
\beq
  R_3 (t) = - \s2th \sin{\lrb{ \fr{\De}{2 E_0} t }}
\quad .
\label{3R1t}
\eeq
Returning to the flavor basis,
we obtain the desired solution:
$$
  r_1 (t) =
   \cos^22\th +
   \sin^22\th \cos{\lrb{ \fr{\De}{2 E_0} t }}
\quad ,
$$
$$
  r_2 (t) =
    - \s2th \c2th
      \lrb{ 1 - \cos{\lrb{ \fr{\De}{2 E_0} t }} }
\quad ,
$$
\beq
  r_3 (t) = - \s2th
       \sin{\lrb{ \fr{\De}{2 E_0} t }}
\quad .
\label{3r1t=}
\eeq

Summarizing,
the solution in the dense neutrino parameter region
is given by
\eqs\bref{3vjt} and \bref{3r1t=}.
These equations
describe self-maintained oscillations.
All neutrinos oscillate in unison.
Note that perfect alignment
is obtained in the limit $\ka^j \to 0$.

When some $\ka^j$ are large,
the corresponding neutrinos do not participate
in the collective mode.
If most neutrinos have small $\ka^j$
then self-maintained coherence still occurs
but with a smaller amplitude.
The criterion for self-maintained coherence
for a pure neutrino gas is
$ \ka_0 < 1$,
where
\beq
    \ka_0 \equiv
   \fr{ \De }{ 2 \sqrt{2} G_F E_0 n_{\nu} }
\quad .
\label{kappa0}
\eeq

We have compared our analytical solution
to numerical simulations.
Excellent agreement is obtained
when all $\ka^j \ll 1$.
Even for the case
in Figure 8
of ref.\ \ct{samuel93},
for which 10\% of the neutrinos had $\ka^j >1$,
agreement between the analytical approach
and numerical simulations
is to about 5\% for the oscillation period
and the amplitude.

An intuitive understanding of alignment
and self-maintained coherence
is as follows.
Assume that most neutrino vectors are aligned.
These vectors point along
the average vector $\vec r (t)$
and collectively rotate around $\vec \De$.
Consider a particular neutrino
with a higher energy than average.
Let $\vec v$ be its vector.
In the absence of the nonlinear term,
$\vec v$ rotates around $\vec \De$
at a relatively slow rate.
Suppose
$\vec v $ begins to lag $\vec r$.
Then, because the nonlinear term is much
bigger than the vacuum term,
the neutrino experiences
a large magnetic field in the direction
of $\vec r (t)$.
Consequently, $\vec v$
rotates around $\vec r$.
After half a period,
$\vec v$ will have rotated to a position
leading the group.
Hence, $\vec v$ cannot lag behind or otherwise separate
from the group.

A similar argument holds
for any neutrino with energy lower than average.
If $\vec v$ begins to lead $\vec r$,
it experiences a large magnetic field
in the direction of $\vec r (t)$ and so rotates
around $\vec r $ rather than $\vec \De$.
It follows that
neutrinos with energies different from the average
do not rotate around $\vec \De$ at varying rates
but instead stay together in a group.
This is alignment.
Since the group follows $\vec r$,
which rotates around $\vec \De$,
oscillatory behavior arises.
This is self-maintained coherence.

\secttit{IV.\ The Neutrino-Antineutrino System}
\setcounter{section}{4}
\setcounter{equation}{0}

In this section,
we study self-maintained coherence
for a dense gas
containing both neutrinos and antineutrinos.
Numerical simulations
reveal that alignment holds separately
for neutrinos and antineutrinos
\ct{ks93,ks94}.
Consequently,
the approximations
\beq
 \vec v^j (t) \approx
  \fr{\expect{\vec v (t)}}{N_{\nu}} \equiv
                 \vec r_{v} (t)
\ , \quad \quad
 \vec w^k (t) \approx
   \fr{\expect{\vec w (t)}}{N_{\bar \nu}} \equiv
                 \vec r_{w} (t)
\quad
\label{4vjt}
\eeq
for the $j$th neutrino and the $k$th antineutrino are good.
By summing over $j$ and $k$
in \eqs\bref{dvecvjdt} and \bref{dvecwkdt},
differential equations
for $\vec r_{v}$ and $\vec r_{w}$ are obtained:
\beq
 {{d\vec r_{v}} \over {dt}} =
    \vec r_{v} \times
    \lrb{ {{\vec \De } \over {2 E_0}} -
           {\vec V_{\nu \nu}} }
\quad ,
\label{4dvecrvdt}
\eeq
\beq
 {{d\vec r_{w}} \over {dt}} =
    \vec r_{w} \times
    \lrb{ {{\vec \De } \over {2 \overline E_0}} +
           {\vec V^{*}_{\nu \nu}}  }
\quad ,
\label{4dvecrwdt}
\eeq
where
\beq
  \vec V_{\nu \nu} = \sqrt{2} G_F
     \lrb{ n_{\nu} \vec r_v - n_{\bar \nu} \vec r_w^{\ *} }
\quad .
\label{4vecVnunu}
\eeq
In \eq\bref{4dvecrvdt},
$1/E_0$ is the average inverse neutrino energy
\bref{31E0},
and $1/\overline E_0$
is the analogous quantity for antineutrinos.

For definiteness,
we consider the situation
with an equal number of
electron neutrinos and antineutrinos
placed in the box at time $t=0$,
$n_{\nu} = n_{\bar \nu}$.
Then, the initial conditions are
\beq
\vec v^j ( 0 ) = \lrb{ 1, 0, 0 }
\ , \quad \quad
\vec w^k ( 0 ) = \lrb{ 1, 0, 0 }
\quad .
\label{4vecvj0}
\eeq
Also, for simplicity we take the antineutrinos
to have the same average inverse energy as neutrinos:
$\overline E_0 = E_0$.
This holds, for example,
in the more restricted case
when the energy distributions
of neutrinos and antineutrinos
are the same,
$\overline E^k = E^j$ for all $k=j$.

The symmetry of the initial conditions
suggests the ansatz
\beq
 \vec r_w (t) = \vec r_v (t)
\quad .
\label{4rw1t}
\eeq
It follows that
\eqs\bref{4dvecrvdt} and \bref{4dvecrwdt}
are equivalent,
so it suffices to solve one of the pair
to demonstrate consistency of the ansatz.
The nonlinear term
in \eq\bref{4vecVnunu}
only has a third component,
$
  \vec V_{\nu \nu} =
    2 \sqrt{2} n_{\nu} G_F r_{v3} \lrb{0,0,1}
$.
The problem therefore reduces to solving
the equations
$$
   {{dr_{v1} } \over {dt}} =
      \lrb{ \fr{\De}{2 E_0} \s2th } r_{v3} -
    2 \sqrt{2} n_{\nu} G_F r_{v2} r_{v3}
\quad ,
$$
$$
   {{dr_{v2} } \over {dt}} =
      \lrb{ \fr{\De}{2 E_0} \c2th } r_{v3} +
    2 \sqrt{2} n_{\nu} G_F r_{v1} r_{v3}
\quad ,
$$
\beq
   {{dr_{v3} } \over {dt}} =
    - \fr{\De}{2 E_0}
      \lrb{ r_{v1} \s2th + r_{v2} \c2th }
\quad
\label{4drv1dt}
\eeq
determining the components of average neutrino vector.

These equations again simplify
in the vacuum-mass-eigenstate basis
$\vec R (t)$ given
in \eq\bref{3R1}.
The vector $\vec R $
obeys \eq\bref{4drv1dt} for $\vec r$
with $\th \to 0$.
In what follows,
it is convenient to make
the further change of variables
\beq
 R_1 (t) = y_1 (s)
\ , \quad \quad
 R_2 (t) = y_2 (s)
\ , \quad \quad
 R_3 (t) = \sqrt{\fr{\ka_0}{2}} y_3 (s)
\quad ,
\label{4R1t}
\eeq
where $\ka_0$ is given
in \eq\bref{kappa0}
and where
\beq
  s = \mu t
\quad ,
\quad
 \mu =
  \fr \De {E_0} \sqrt{\fr 1 {2 \ka_0}}
\quad .
\label{4s}
\eeq
The oscillation equations simplify to
\beq
   {{dy_{1} } \over {ds}} =
      - y_{2} y_{3}
\quad ,
\label{4dy1dt}
\eeq
\beq
   {{dy_{2} } \over {ds}} =
        y_{1} y_{3} + \fr{\ka_0}{2} y_{3}
\quad ,
\label{4dy2dt}
\eeq
\beq
   {{dy_{3} } \over {ds}} =
      - y_{2}
\quad .
\label{4dy3dt}
\eeq
The initial conditions
\bref{4vecvj0} become
\beq
\vec R ( 0 ) = \vec y ( 0 ) = \lrb{ \c2th, \s2th, 0 }
\quad .
\label{4vecR0}
\eeq

To proceed,
we can take advantage of
the conservation of neutrino number,
which in the present variables is expressed as
\beq
  y_1^2 + y_2^2 + \fr{\ka_0}{2} y_3^2 = 1
\quad .
\label{4y12y22kappa02y32}
\eeq
This equation,
which is a consequence
of \eqs\bref{4dy1dt}--\bref{4vecR0},
specifies $y_2$ in terms of $y_1$ and $y_3$.
Furthermore,
an equation for $y_3$ in terms of $y_1$
is obtained by substituting
\bref{4dy3dt} into \bref{4dy1dt}
and integrating.
The above observations
determine $y_2$ and $y_3$
in terms of $y_1$ as
\beq
  y_2 =
   \pm \sqrt{ 1 - y_1^2 -
   \ka_0 \left( { y_1 - \c2th } \right)
            }
\quad ,
\quad
  y_3 = \pm \sqrt{
     2\left( { y_1 - \c2th } \right)
                 }
\quad .
\label{4y3}
\eeq
Specifying the signs corresponds to
specifying different stages of the motion,
as discussed below
(see \eq\bref{4stages}).

At this point,
we need only obtain $y_1 (t)$.
A differential equation for this variable
can be found by
differentiating \bref{4dy1dt}
with respect to $s$,
using
\eqs\bref{4dy2dt} and \bref{4dy3dt},
incorporating
\eqs\bref{4y12y22kappa02y32} and \bref{4y3},
multiplying by $dy_1 /ds$,
and integrating once.
The integration constant
is fixed using
\eq\bref{4vecR0}.
This procedure gives
\beq
  \fr{1}{2} {\lrb{{{dy_{1} } \over {ds}}}}^2 =
     - {V} \lrb{y_1}
\quad ,
\label{412dy1dw2}
\eeq
where
\beq
 -{V} \left( {y_1} \right) =
    \left( { y_1 - \c2th } \right)
    \left( { 1 - y_1^2 - \ka_0
      \left( {y_1 - \c2th } \right)
           } \right)
\quad .
\label{4Vy1}
\eeq
Equation \rf{412dy1dw2}
is analogous to one
describing the motion of a particle of unit mass
moving in the potential ${V} \lrb{y_1}$
associated to an anharmonic oscillator with a cubic term.

The motion of $y_1$ is between
$y_1^{\min}$ and $y_1^{\max}$,
where
\beq
  y_1^{\min} = \c2th
\ , \quad \quad
  y_1^{\max} = x_+ + \c2th
\quad .
\label{4y1min}
\eeq
The point $x_+$ determines one of the two zeros
of the potential
$-{V} \left( {x + \c2th } \right)\equiv
          - x \left( {x - x_+ } \right)
          \left( {x   + x_- } \right) $,
where
\beq
  x_\pm \equiv
   \sqrt{ 1 + \ka_0 \c2th + \ka_0^2/4 } \mp
     \c2th \mp \ka_0/2
\quad .
\label{4x+}
\eeq
In the region
$ y_1^{\min} \le y_1 \le y_1^{\max}$,
it follows that
$ - {V} \left( {y_1} \right) \ge 0$,
so both sides
of \eq\bref{412dy1dw2}
are positive.

The solution
of \eq\bref{412dy1dw2}
is by quadrature in $x$.
Changing the integration variable using
$x = x_+ w^2$
gives the implicit solution
\beq
  s =
    \sqrt{ {2 \over {x_-}} }
  \int_0^{
   \sqrt{{{y_1\left( s \right) - \c2th } \over {x_+}}}}
   {{dw} \over {
    \sqrt{\left( {1 -     w^2} \right)
          \left( {1 + q^2 w^2} \right)
         }     }
   }
\quad ,
\label{4s=b}
\eeq
where
$q^2 = x_+ / x_-$.
This expression can be inverted for $y_1(s)$
using the sine-amplitude and delta-amplitude
Jacobi elliptic functions
${\rm sn}$ and ${\rm dn}$,
defined as
\beq
 u =
    \int_0^{{\rm sn}\lrb{u,k}}
    \fr{dt}{\sqrt{\lrb{1 - t^2} \lrb{1- k^2 t^2}}}
\quad ,
\quad
 v =
    \int_0^{{\rm dn}\lrb{v,k}}
    \fr{dt}{\sqrt{\lrb{1 - t^2}
         \lrb{1 - \lrb{1 - k^2} t^2}}
           }
\quad .
\label{4u}
\eeq
We finally obtain
\beq
  y_1 \left( s \right) =
   \c2th +
  {
     { \sin^2{ \lrb{2\th} } }
    \over
     { x_{0}}
  }
   {
     { {\rm sn}^2 \left( {\sqrt{{{x_{0}} \over 2}} s ,
              \sqrt{{{x_+} \over {x_{0}}}}} \right)}
    \over
     { {\rm dn}^2 \left( {\sqrt{{{x_{0}} \over 2}} s ,
              \sqrt{{{x_+} \over {x_{0}}}}} \right)}
   }
\quad ,
\label{4y1s}
\eeq
where
$
  x_{0} \equiv x_+ + x_- =
   2 \sqrt{1 + \ka_0 \c2th + \ka_0^2/4}
$.

The motion consists of four stages:
\beq
\matrix{
  {\rm stage\ 1:} \quad &
      y_2  \ge 0 \ ,&      y_3  \le 0 \ ,&
 \dds{y_1} \ge 0 \ ,& \dds{y_2} \le 0 \ ,&
 \dds{y_3} \le 0 \ , \quad \cr
  {\rm stage\ 2:} \quad &
      y_2  \le 0 \ ,&      y_3  \le 0 \ ,&
 \dds{y_1} \le 0 \ ,& \dds{y_2} \le 0 \ ,&
 \dds{y_3} \ge 0 \ , \quad \cr
  {\rm stage\ 3:} \quad &
      y_2  \le 0 \ ,&      y_3  \ge 0 \ ,&
 \dds{y_1} \ge 0 \ ,& \dds{y_2} \ge 0 \ ,&
 \dds{y_3} \ge 0 \ , \quad \cr
  {\rm stage\ 4:} \quad &
      y_2  \ge 0 \ ,&      y_3  \ge 0 \ ,&
 \dds{y_1} \le 0 \ ,& \dds{y_2} \ge 0 \ ,&
 \dds{y_3} \le 0 \quad . \cr
       }
\quad
\label{4stages}
\eeq
Since $y_1 \ge \c2th$,
it follows that $y_1 > 0$ for all stages of the motion.
During the motion,
$y_2^{min} \le y_2 \le y_2^{max}$,
where $y_2^{max} = \s2th = - y_2^{min} $.

The motion of $\vec y$ is roughly
circular about $\vec \De$.
Recall that
$\vec \De$ points along the $1$-axis
of the vacuum-mass-eigenstate basis.
At $t=0$,
$\vec y = \lrb{\c2th,\s2th,0}
= \lrb{y_1^{min},y_2^{max},0}$,
so $\vec y$ points in the direction
of the $1$-axis of the flavor basis.
During stage 1,
$\vec y$ drops below the $1$-$2$ plane.
It then passes to stage 2 when it drops below $\vec \De$.
At the 1-to-2 transition point,
$y_1 = y_1^{max}$ and $y_2 = 0$.
The vector $\vec y$ continues its motion
below the $1$-$2$ plane during stage 2,
until $y_1$ and $y_2$ obtain their minimum values
and $y_3 = 0$.
In stages 3 and 4,
the motion is reversed,
except that $y_3$ is positive so that
$\vec y$ is above the $1$-$2$ plane.
The maximum value of $y_1$ is again achieved
at the 3-to-4 transition,
where $\vec y$ is above $\vec \De$
and $y_2 = 0$.
The cycle is completed when
$\vec y$ returns to
$\vec y = \lrb{\c2th,\s2th,0}$.
Hence,
one entire cycle of motion
involves two cycles of $y_1$.
The signs of $y_2$ and $y_3$
in \eqs\bref{4y3}
are determined
from the second and third columns
in \eq\bref{4stages}.

Figures 1a-c display the behavior
over two periods of each of the three components
of the vector $\vec r$
for the case with $\sin^2 2\th = 0.81$
and $\ka_0 = 0.1$.
The time scale is plotted in terms of the $s$ variable,
which is equivalent to the time $t$ measured in units
of $1/\mu$.
It is convenient to use $s$ so that $\De$ and $E_0$
need not be specified
(compare with \eq\bref{4s}).
Extra oscillations appear in the second component,
plotted in Fig.\ 1b.
They arise from the projection of the orbit onto the 2 axis.
The variables $y_1$, $y_2$, $y_3$
have no such effects.
Figure 2 shows the same orbit in a three-dimensional
plot.

In general,
the half-period of $y_1$ in $s$
is determined from
\eq\bref{4s=b}
by setting $y_1 = y_1^{max}$.
Since a complete cycle involves
two $y_1$ cycles,
one obtains
\beq
  S_{\nu \bar \nu} = 4 \sqrt {{2 \over {x_-}}}
   \int_0^1
  { {dw}
  \over
    {\sqrt{\left( {1 -     w^2} \right)
           \left( {1 + q^2 w^2} \right)}
    }
  }
\quad
\label{4Snubarnu}
\eeq
for the period $S_{\nu \bar \nu}$ in $s$.
Equation \bref{4s} then implies that
the period $T_{\nu \bar \nu}$ in $t$ is
\beq
  T_{\nu \bar \nu} =  \fr{S_{\nu \bar \nu}}{\mu}
\quad .
\label{4Tnubarnu}
\eeq
Hence, the period $T_{\nu \bar \nu}$ is of the order
of the geometric mean of the time scales
associated with the vacuum and nonlinear terms:
$T_{\nu \bar \nu} \sim \sqrt{T_{\De}{T_{\nu}}}$.
The motion for the neutrino-antineutrino gas is
thus on the order of $1/\sqrt{\ka_0}$ times faster
than in the pure neutrino case of Sect.\ 3.

Anther interesting feature of
the behavior of the neutrino-antineutrino system
is its near planarity,
apparent since $R_3$ is related to $y_3$
by a factor of $\sqrt{{\ka_0}/{2}}$.
The range of $R_3$ is of order
$\sqrt{\ka_0} \ll 1$.
Hence,
the bulk of the motion is in the $1$-$2$ plane.
Planarity is a feature observed
in the numerical simulations
of refs.\ct{kps93,ks93,ks94}.
In the variables $\vec R$ or $\vec r$,
the orbit is similar to a highly eccentric ellipse.

\secttit{V.\ Summary}
\setcounter{section}{5}
\setcounter{equation}{0}

In this paper,
we have provided analytical solutions to
the nonlinear equations describing
the behavior of a gases
containing two flavors of neutrinos,
both with and without antineutrinos.
For a dense pure neutrino gas,
the solution is given by
\eqs\bref{3vjt} and \bref{3r1t=},
while for
a dense neutrino-antineutrino gas
the solution is specified by
\eqs\bref{3R1}, \bref{4R1t}, \bref{4s},
\bref{4y3}, \bref{4y1s} and \bref{4stages}.

The behavior
of the neutrino-antineutrino gas
differs from that of the pure neutrino case.
The former is controlled by elliptic functions,
whereas the latter is governed by
trigonometric functions.
Our analytical results agree in detail with
prior numerical simulations
in the region with $\ka \ll 1$.

We have demonstrated analytically that
self-maintained coherent flavor oscillations
occur when the gas density
exceeds a critical value,
given in terms of the mean neutrino energy
and the neutrino masses.
Oscillations of this type may have occurred
in the early Universe.

\noindent
\secttit{Acknowledgements}

We thank J.\ Pantaleone for discussions.
This work is supported in part
by the United States Department of Energy
(grant numbers DE-FG02-91ER40661 and DE-FG02-92ER40698),
by the Alexander von Humboldt Foundation,
and by the PSC Board of Higher Education at CUNY.

\newpage
{\bf\large\noindent References}
\vglue 0.4cm

\def\plb #1 #2 #3 {Phys.\ Lett.\ B #1 (19#2) #3.}
\def\mpl #1 #2 #3 {Mod.\ Phys.\ Lett.\ A #1 (19#2) #3.}
\def\prl #1 #2 #3 {Phys.\ Rev.\ Lett.\ #1 (19#2) #3.}
\def\pr #1 #2 #3 {Phys.\ Rev.\ #1 (19#2) #3.}
\def\prd #1 #2 #3 {Phys.\ Rev.\ D #1 (19#2) #3.}
\def\npb #1 #2 #3 {Nucl.\ Phys.\ B#1 (19#2) #3.}
\def\ptp #1 #2 #3 {Prog.\ Theor.\ Phys.\ #1 (19#2) #3.}
\def\jmp #1 #2 #3 {J.\ Math.\ Phys.\ #1 (19#2) #3.}
\def\nat #1 #2 #3 {Nature #1 (19#2) #3.}
\def\prs #1 #2 #3 {Proc.\ Roy.\ Soc.\ (Lon.) A #1 (19#2) #3.}
\def\ajp #1 #2 #3 {Am.\ J.\ Phys.\ #1 (19#2) #3.}
\def\lnc #1 #2 #3 {Lett.\ Nuov.\ Cim. #1 (19#2) #3.}
\def\nc #1 #2 #3 {Nuov.\ Cim.\ A#1 (19#2) #3.}
\def\jpsj #1 #2 #3 {J.\ Phys.\ Soc.\ Japan #1 (19#2) #3.}
\def\ant #1 #2 #3 {At. Dat. Nucl. Dat. Tables #1 (19#2) #3.}
\def\nim #1 #2 #3 {Nucl.\ Instr.\ Meth.\ B#1 (19#2) #3.}

\newpage

\noindent
FIGURE CAPTIONS

\medskip

Figure 1. Components of the vector $\vec r$
as a function of scaled time, $s=\mu t$,
for the case $\sin^2 2\th = 0.81$ and $\ka_0 = 0.1$.
(a) The component $r_1$.
(b) The component $r_2$.
(c) The component $r_3$.

\medskip

Figure 2. The three-dimensional orbit
for the case $\sin^2 2\th = 0.81$ and $\ka_0 = 0.1$.

\vfill
\end{document}